\documentclass[floats,floatfix,showpacs,amssymb,prd,twocolumn,superscriptaddress,nofootinbib,longbibliography]{revtex4-1}

\usepackage{anyfontsize}
\usepackage{amssymb}
\usepackage{amsmath}
\usepackage{amsfonts}
\usepackage{amsbsy}
\usepackage[utf8]{inputenc} 
\usepackage{newtxtext}
\usepackage{newtxmath}
\usepackage{lipsum,graphicx,afterpage}
\usepackage{tensor}
\usepackage{mathrsfs}

\usepackage{bm}

\usepackage[utf8]{inputenc}

\usepackage{graphicx}
\usepackage{epsfig}
\usepackage{epstopdf}

\usepackage[usenames,dvipsnames]{xcolor}

\usepackage{verbatim,mathtools,needspace,enumitem,etoolbox}

\definecolor{linkcolor}{rgb}{0.0,0.3,0.5}

\usepackage[unicode, colorlinks=true, linkcolor=linkcolor, citecolor=linkcolor, filecolor=linkcolor,urlcolor=linkcolor, pdfusetitle]{hyperref}

\usepackage{epstopdf}

\usepackage{bm}
\usepackage{dcolumn}

\usepackage{longtable}
\usepackage{enumerate}
 \usepackage[normalem]{ulem}

\hypersetup{colorlinks=true,
            citecolor=blue,
            linkcolor=blue,
            urlcolor=blue}
\usepackage[T1]{fontenc}
\usepackage{subfigure}
\usepackage{amsmath}
\usepackage{float}
\usepackage{mathrsfs}
\usepackage[dvipsnames]{xcolor}
\usepackage{soul}

\definecolor{darkgreen}{RGB}{1,212,57}

\setcitestyle{numbers,square}

\setcitestyle{numbers,square}

\begin{document}
\title{Tidal forces in the charged Hayward black hole spacetime
}

\author{Haroldo C. D. Lima Junior}
\email{haroldo.ufpa@gmail.com} 
\affiliation{Faculdade de F\'{\i}sica,
Universidade Federal do Par\'a, 66075-110, Bel\'em, PA, Brasil }

\author{Lu\'{\i}s C. B. Crispino}
\email{crispino@ufpa.br} 
\affiliation{Faculdade de F\'{\i}sica,
Universidade Federal do Par\'a, 66075-110, Bel\'em, PA, Brasil }

\date{\today}

\begin{abstract}
Tidal forces produced by black holes are an important result of General Relativity, related to the spacetime curvature tensor. Among the astrophysical implications of tidal forces, stands out tidal disruption events. We analyze the tidal forces in the spacetime of  an electrically charged Hayward regular black hole, obtaining the components of the tidal tensor and the geodesic deviation equation.  We find that the radial and angular tidal forces may vanish and change sign, unlike in the Schwarzschild spacetime. We note that tidal forces are finite at the origin of the radial coordinate in this regular black hole spacetime. We obtain the geodesic deviation vector for a body constituted of dust infalling towards the black hole with two different initial conditions.
\keywords{Tidal Forces, Regular Black Holes, General Relativity.}
\end{abstract}

\maketitle

\section{Introduction}
Black holes (BHs) have been thought to be the best candidates for supermassive objects present in the center of active galaxies \cite{Narayan}. This has been confirmed in the recent observation by the Event Horizon Telescope collaboration (EHT) of the shadow of a BH in the center of Messier 87 (M87) galaxy \cite{EHT}. The earliest known solution of Einstein's equations is the Schwarzschild geometry, which represents the spacetime outside a spherically symmetric and chargeless nonrotating body. The Schwarzschild solution can also describe the spacetime around a  nonrotating and electrically uncharged BH \cite{chandrasekhar}. Another solution of the Einstein's equations is the Reissner-Nordstr{\"o}m (RN) geometry, which represents the spacetime around an electrically charged, nonrotating and spherically symmetric BH \cite{chandrasekhar}.  Besides Schwarzschild and RN BH solutions, we also have the axisymmetric exact solutions of Einstein's equations.  The Kerr solution is an axisymmetric geometry that represents the spacetime outside an electrically uncharged and rotating BH \cite{kerr}. The Kerr-Newman solution is also an axisymmetric geometry that represents the spacetime outside a rotating BH endowed with electric charge \cite{kerrnewman}. BHs possess remarkable physical properties, which lead to interesting phenomenology as, for instance, absorption \cite{ACS:1977, ACS:2005, ACS:2008, ACS:2009, ACS:2010, ACS:2011, Benone:2014qaa, Benone:2016, ACS: 2017}, scattering \cite{SC: 2006, SC: 2014, SC:2015, SC:2015.1}, and radiation emission~\cite{R_EM,Castineiras:2005ww, Crispino:2008zza,R_EM1,R_EM2}.

All the BH solutions mentioned above posses a curvature singularity, where geometrical and physical quantities diverge. The existence of BH solutions free of curvature singularities was pointed out by 
J.~M.~Bardeen~\cite{bardeen}, who proposed a regular BH solution. This Bardeen BH is an exact solution of Einstein's equations coupled to nonlinear electrodynamics
 \cite{BardeenNED}. 

The Hayward BH~\cite{Hayward} is also an important example of regular BH 
 with properties investigated in the literature as, for instance, quasinormal modes~\cite{HaywardQNM,HaywardQNM1,HaywardQNM2} and  geodesics~\cite{Haywardgeo}, and its rotating generalizations have also been  found~\cite{Rot_Hayward}.  More recently, Frolov proposed a generalization of the Hayward BH including electric charge~\cite{Frolov::PRD2016}, from which it is possible to recover the RN line element by taking a suitable limit. A rotating generalization of the charged Hayward BH and its shadows was studied in 
Ref.~\cite{Shadow_CHayward}.
  
It is a well-known fact that radial tidal forces in a \sloppy Schwarzschild spacetime stretch a body falling towards the event horizon, meanwhile the same body is compressed in the angular directions \cite{dinverno, hobson}. In the last few years tidal forces were also analyzed for RN BH~\cite{RN_TF}, Kiselev~\cite{Kiselev_TF} and other regular BHs~\cite{RBH_TF}. The tidal forces in Kerr spacetime were analyzed, for instance, in Refs.~\cite{Marck:1983,AxisTF,R2,R3,R4,Junior:2020yxg}. Tidal forces play an important role in astrophysics. For instance, a star may be disrupted due to the tidal forces produced by a BH \cite{TDE1}, what is known as a tidal disruption event (TDE). Such TDEs may power bright flares as x-ray~\cite{TDE2}, ultraviolet~\cite{TDE3} and optical~\cite{TDE4} radiation.

We study the tidal forces produced in an electrically charged Hayward BH spacetime. We find the components of the tidal tensor and present the geodesic deviation equation. We also find the solutions for the geodesic deviation vector in the radial and angular directions for a neutral body, constituted of dust, infalling radially towards a charged Hayward BH.

The remaining of this paper is structured as follows. In Sec.~\ref{CH_ST}, we review the charged Hayward BH spacetime and its properties. We investigate the radial geodesics in such spacetime in Sec.~\ref{Radial-geo} and  the tidal forces in Sec.~\ref{TF_Hayward}. In Sec.~\ref{dev_vec_sol}, we obtain the geodesic deviation vector in the charged Hayward BH spacetime and in Sec.~\ref{Conclusion} we present our conclusions. We use the metric signature ($-$,$+$,$+$,$+$) and set the speed of light ($c$) and the Newtonian gravitational constant ($G$) equal to the unity.

\section{The charged Hayward BH Spacetime}  
\label{CH_ST}  
The electrically charged Hayward BH, proposed by Frolov in Ref.~\cite{Frolov::PRD2016}, is described by the line element
\begin{equation}
\label{line_el}ds^2=g_{\mu\nu}dx^\mu\,dx^\nu=-f(r)\,dt^2+\frac{1}{f(r)}\,dr^2+r^2\,d\Omega^2,
\end{equation}
with
\begin{align}
&d\Omega^2\equiv \sin^2\theta\,d\phi^2+d\theta^2,
\end{align}
and
\begin{align}
\label{f_r}&f(r)\equiv 1-\frac{\left(2\,M\,r-Q^2\right)\,r^2}{r^4+(2Mr+Q^2)\,l^2},
\end{align}
where $g_{\mu\nu}$ are the covariant components of the metric tensor, $l$ is the parameter present in the Hayward solution~\cite{Hayward}, $M$ is the mass of the BH and $Q$ is its electric charge. In the limit $l\rightarrow0$, the line element \eqref{line_el} reduces to the RN spacetime. On the other hand, for $Q\rightarrow0$, it reduces to the Hayward regular BH. The Schwarzschild line element is obtained when both limits $Q\rightarrow 0$ and $l\rightarrow 0$ are taken  in Eq.~\eqref{line_el}. Next to the origin of the radial coordinate ($r\approx 0$), we have
\begin{equation}
\left.\label{metric_r_0}f(r)\right|_{r\approx 0}\approx 1+\frac{r^2}{l^2}+\mathcal{O}(r^3),
\end{equation} 
from what we conclude that the metric tensor components are regular at the origin ($r=0$) if $l\neq 0$. On the other hand, very far away from the BH ($r \approx \infty$), we have
\begin{equation}
\label{metric_r_infty}\left. f(r)\right|_{r\approx \infty}\approx 1-\frac{2\,M}{r}+\frac{Q^2}{r^2}+\mathcal{O}(r^{-4}).
\end{equation}
From Eq.~\eqref{metric_r_infty} we note that the charged Hayward BH solution behaves as the RN solution for large values of the radial coordinate, and from Eq.~\eqref{metric_r_0} we conclude that next to the origin it behaves like the de Sitter solution.

\begin{center}
\begin{figure}[h!]
\center
\includegraphics[scale=0.7]{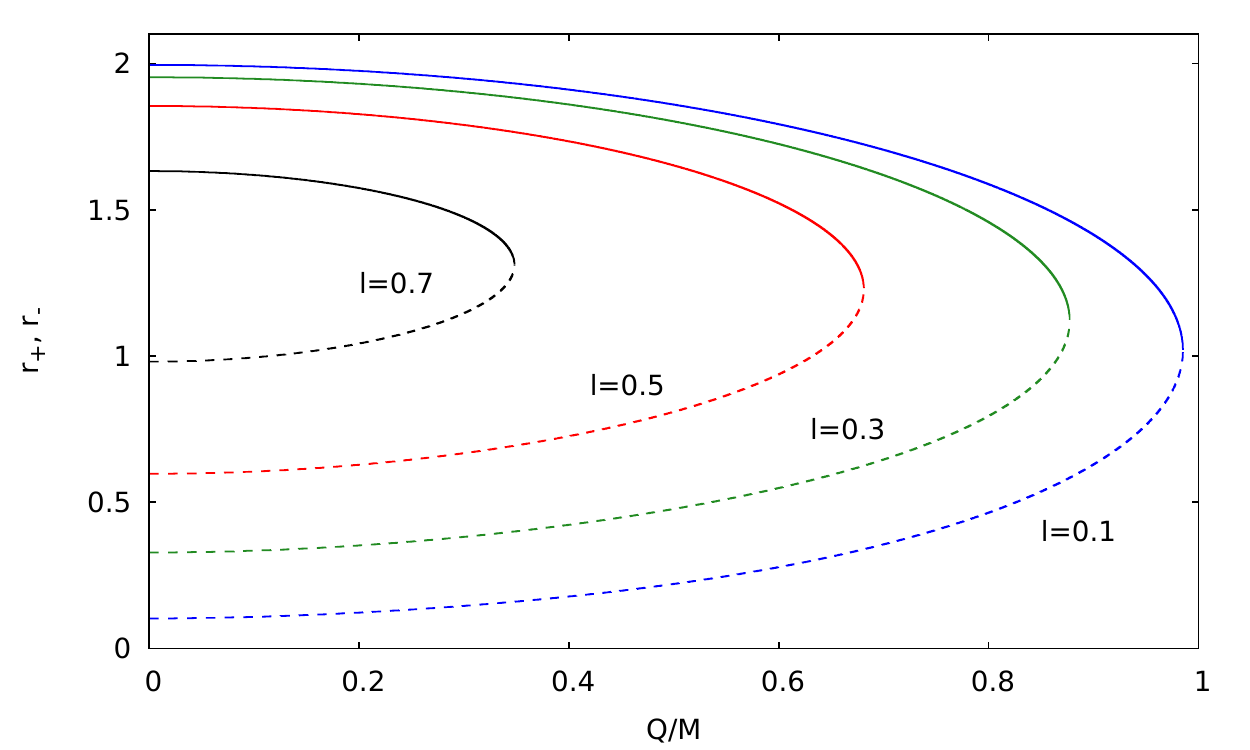}
\caption{Horizons of the charged Hayward BH spacetime described by the line element \eqref{line_el}, in terms of $Q/M$, for different values of $l$. The radial coordinates of the event horizon, $r_+$, are represented by solid lines, whereas the radial coordinates of the Cauchy horizon, $r_-$, are represented by dashed lines. }
\label{horizons}
\end{figure}
\end{center}

The horizons of the charged Hayward spacetime are found imposing
\begin{equation} 
\label{horizon_eq}g^{rr}=f(r)=0.
\end{equation} 
Eq.~\eqref{horizon_eq} may have two real and positive solutions, one of which (the outer one, $r_+$) is the event horizon and the other (the inner one, $r_-$) is the Cauchy horizon. In Fig.~\ref{horizons}, we plot the radial coordinate of the event horizon and of the Cauchy horizon in terms of the electric charge $Q$, for different choices of the parameter $l$. We note that the event horizon radial coordinate, $r=r_+$, decreases as we increase the values of $Q$ and $l$. We also see that, the Cauchy horizon radial coordinate, $r=r_-$, increases as we increase the values of $Q$ and $l$. For each fixed value of $l$, we may find an extreme charged Hayward spacetime, where the event horizon and the Cauchy horizon radial coordinates coincide ($r_+=r_-$), as it can be seen in Fig.~\ref{horizons}. In this case, we have
\begin{equation}
f(r)=\frac{df(r)}{dr}=0.
\end{equation} 
The relation between the mass and the charge for extreme charged Hayward BHs is given in a parametric form as~\cite{Frolov::PRD2016}:
\begin{eqnarray}
Q^2=\frac{r^4\left(r^2-3l^2\right)}{r^4+4l^2r^2-l^4},\\
M=\frac{\left(r^2+2l^2\right)r^3}{r^4+4l^2r^2-l^4}.
\end{eqnarray}
As pointed out in Ref.~\cite{Frolov::PRD2016}, for small values of $Q$, we have
\begin{equation}
M\approx \frac{3\sqrt{3}}{4}l+\frac{1}{\sqrt{3}}\frac{Q^2}{l}+\mathcal{O}(Q^4),
\end{equation}
and for $Q=0$, we have $M=\frac{3\sqrt{3}}{4}l$, which is the relation for an extreme chargeless Hayward BH.

\section{Radial geodesics}
\label{Radial-geo}

Let us investigate radial geodesics of massive particles in charged Hayward BH spacetimes. From Eq.~\eqref{line_el}, we obtain
\begin{equation}
\label{radial_geo}-1= -f(r)\,\dot{t}^2+f^{-1}(r)\,\dot{r}^2,
\end{equation}
where the overdots represent differentiation with respect to the particle's proper time $\tau$. Since we are interested in radial motion, we have set $\dot{\theta}=\dot{\phi}=0$. Due to the existence of one timelike Killing vector and one spacelike Killing vector, we have two conserved quantities along the geodesics, associated, respectively, to the energy and angular momentum of the particle~\cite{Wald}, given by
 \begin{align}
&\label{Energy} E=f(r)\,\dot{t},\\
&\label{Ang_mom}L=r^2\,\sin^2\theta\,\dot{\phi}=0.
 \end{align}
 $E$ is the energy per unit of  rest mass of the particle on a geodesic motion. $L$ is the angular momentum per unit mass of the particle, which is equal to zero, once that we are assuming that the motion is radial. Inserting Eq.~\eqref{Energy} into Eq.~\eqref{radial_geo}, we obtain an energy balance equation, namely:
 \begin{equation}
 \label{radialvel}\dot{r}^2+f(r)=E^2.
 \end{equation}
Assuming that a massive particle is released from rest at the radial position $r=b$, its energy $E$ per unit rest mass is given by 
\begin{equation}
E=\sqrt{1-\frac{\left(2\,M\,b-Q^2\right)\,b^2}{b^4+(2Mb+Q^2)\,l^2}}.
\end{equation} 
 
Defining a ``Newtonian radial acceleration'' \cite{Marion}, as
 \begin{equation}
 \label{newacc}A^{(N)}\equiv\ddot{r},
 \end{equation}
 we find, from Eqs.~\eqref{f_r}, \eqref{radialvel} and \eqref{newacc}, that
 \begin{equation}
\label{Newton_acc} A^{(N)}=\frac{r^5\left(Q^2-Mr\right)+l^2r\left(4M^2r^2+2MQ^2r-Q^4\right)}{\left(r^4+l^2\left(Q^2+2Mr\right)\right)^2}.
 \end{equation}
 There is not a classical analogue for Eq. \eqref{Newton_acc}. However, for $Q=0$ and $l=0$, it is possible to recover an expression similar to the Newtonian one, namely
 \begin{equation}
 A^{(N)}=-\frac{M}{r^2}.
 \end{equation} 
 
\begin{figure}
  \centering
  \subfigure{\includegraphics[scale=0.7]{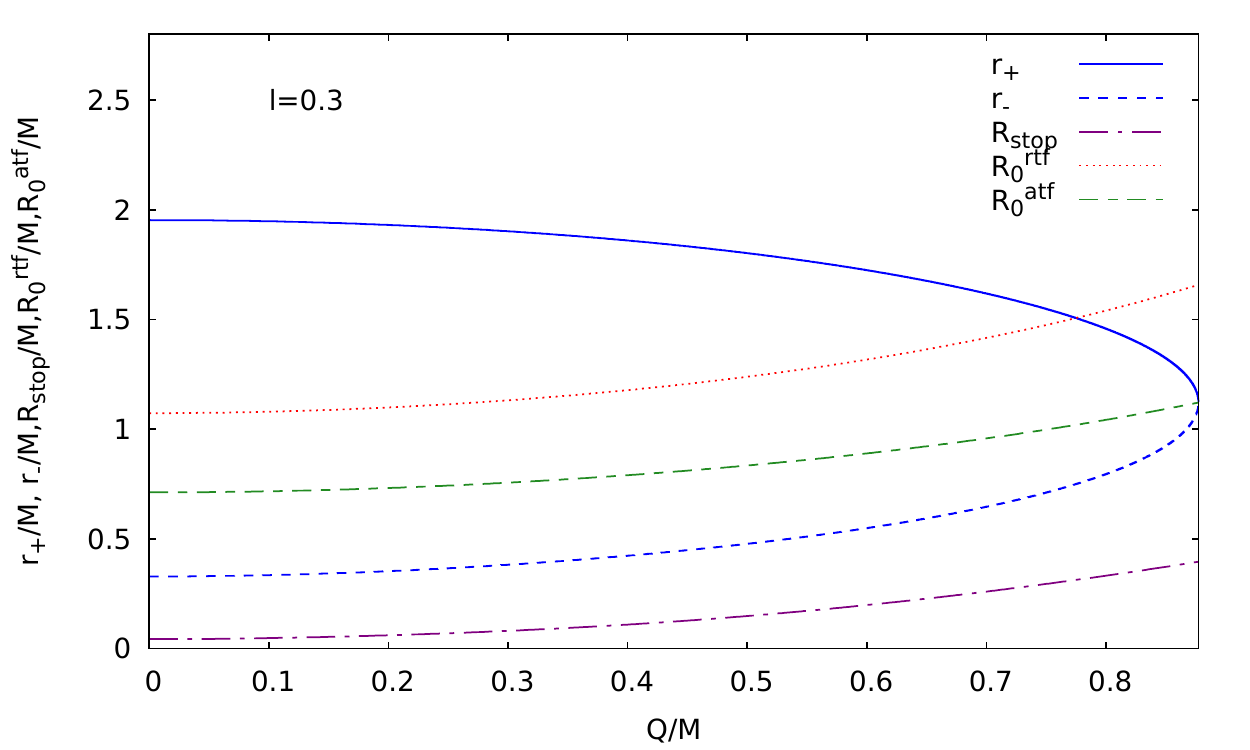}}
  \\
\subfigure{\includegraphics[scale=0.7]{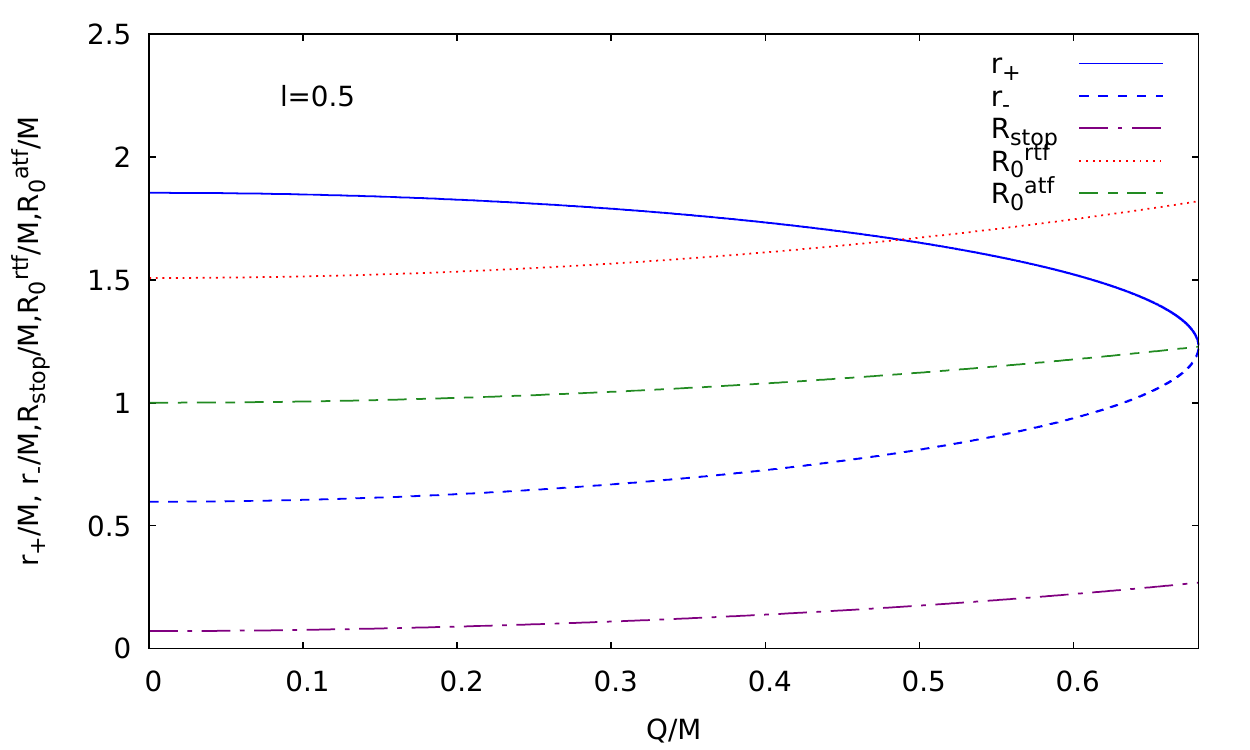}}\\
  \subfigure{\includegraphics[scale=0.7]{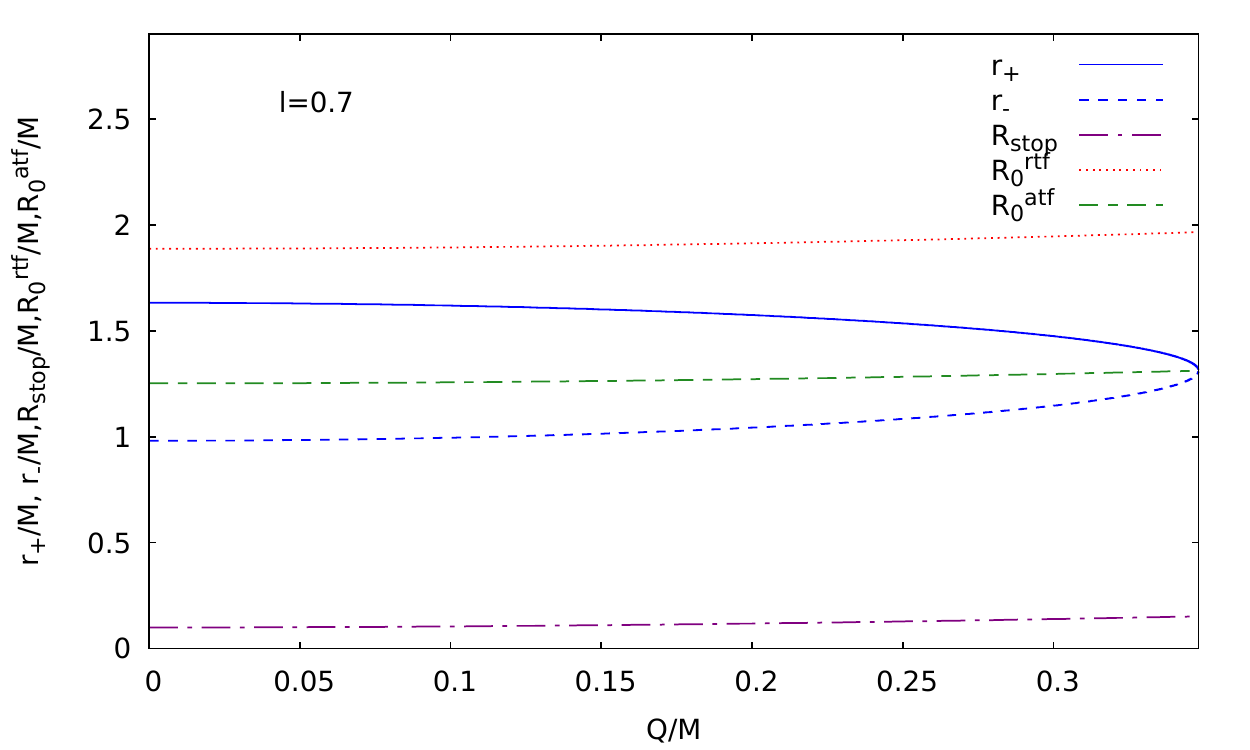}}
\caption{Event horizon $r_+$ (solid blue lines), Cauchy horizon $r_-$ (dashed blue lines), the turning point $R_\text{stop}$ (dotted purple lines), the outermost vanishing radial tidal force $R_0 ^{\ rtf}$ (dotted red lines) and the outermost vanishing angular tidal forces $R_0 ^{\ atf}$ (dotted green lines) as a function of the electric charge $Q$. We have chosen $b=100M$ and different values of $l$. We note that $R_\text{stop}$ is always inside the Cauchy horizon.}
\label{R_stop_Fig}
\end{figure}

A massive particle released from rest at $r=b$ has a turning point at $r=R_{\text{stop}}$, which may be determined by finding the roots of
\begin{equation}
E^2-f(R_{\text{stop}})=0.
\end{equation}
 The turning point $R_{\text{stop}}$ is the location where the particle bounces back in its radial motion and, as it can be seen in Fig.~\ref{R_stop_Fig}, it is always located inside the Cauchy horizon. It is worth noting that the Cauchy horizon is generally associated to instabilities, connected to the so-called Israel-Poisson mass inflation
 (cf. Refs.~\cite{Mass_infl,Mass_infl1}). 

\section{Tidal forces on  a neutral body following radial geodesics}
\label{TF_Hayward}
We now study the geodesic deviation equation in the charged Hayward BH spacetime, following the approach adopted in Ref.~\cite{RN_TF}. We denote as $\eta^\mu$ a spacelike vector which represents the distance between two infinitesimally close particles following geodesics, so that the equation for their relative acceleration is \cite{dinverno}
\begin{equation}
\label{tidalacele}\frac{D^2 \eta^{\hat{\alpha}}}{D\,\tau^2}=K^{\hat{\alpha}}_{\ \ \hat{\beta}}\eta^{\hat{\beta}},
\end{equation} 
where 
\begin{equation}
\label{tidalt}K^{\hat{\alpha}}_{\ \ \hat{\beta}}=R^a_{\ \ b\,c\,d}\,e^{\hat{\alpha}}_{\ a}\,e^{b}_{\ \hat{0}}\,e_{\ \hat{0}}^{c}\,e_{\ \hat{\beta}}^{d}
\end{equation}
is the so-called tidal tensor. The indices with hat are tetrad basis indices, while the indices without hat are coordinate basis indices. For the charged Hayward BH spacetime, we may choose the tetrad basis for a reference frame following a radial geodesic as
\begin{eqnarray}
\label{tetrad0}&&e^\mu_{\ \hat{0}}=\left(\frac{E}{f(r)}, -\sqrt{E^2-f(r)}, 0, 0\right),\\
\label{tetrad1}&&e^\mu_{\ \hat{1}}=\left(-\frac{\sqrt{E^2-f(r)}}{f(r)}, E, 0, 0\right),\\
\label{tetrad2}&&e^\mu_{\ \hat{2}}=\left(0, 0, r^{-1}, 0\right),\\
\label{tetrad3}&&e^\mu_{\ \hat{3}}=\left(0, 0, 0, (r\sin\theta)^{-1}\right).
\end{eqnarray}
We point out that the vectors $e_{\ \hat{\alpha}}^{\mu}$ satisfy the following orthonormality condition
\begin{align}
e_{\ \hat{\alpha}}^{\mu}\,e_{\ \hat{\beta}}^{\nu}\,g_{\mu\,\nu}=n_{\hat{\alpha}\hat{\beta}},
\end{align}
where $n_{\hat{\alpha}\hat{\beta}}$ are the covariant components of the Minkowski metric. Moreover, we have $e^\mu_{\ \hat{0}}=\dot{x}^\mu$.

Computing the components of the Riemann tensor~\cite{chandrasekhar} for the charged Hayward BH spacetime and using Eqs.~\eqref{tidalt}-\eqref{tetrad3}, we find that the nonvanishing tidal tensor components are given by
\begin{align}
\nonumber&K_{\hat{r}\hat{r}}=\frac{1}{\left(r^4+2l^2Mr+Q^2l^2\right)^3}\left[6Q^4\left(l^4Mr+2l^2r^4\right)-Q^6l^4\right.\\
\nonumber & +2Mr^3\left(4l^4M^2-14l^2Mr^3+r^6\right)\\
&\label{Krr} \left.-3Q^2r^2\left(r^6
-4l^4M^2+4l^2Mr^3\right)\right],\\
\label{Ktt}&K_{\hat{\theta}\hat{\theta}}=-\frac{\left[Q^4l^2+Mr^2\left(r^3-4l^2M\right)-Q^2\left(2l^2Mr+r^4\right)\right]}{\left(r^4+2l^2Mr+Q^2l^2\right)^2},\\
\label{Kpp}&K_{\hat{\phi}\hat{\phi}}=-\frac{\left[Q^4l^2+Mr^2\left(r^3-4l^2M\right)-Q^2\left(2l^2Mr+r^4\right)\right]}{\left(r^4+2l^2Mr+Q^2l^2\right)^2}.
\end{align}
Using Eqs.~\eqref{Krr}-\eqref{Kpp} in Eq.~\eqref{tidalacele}, we obtain the relative acceleration between two nearby particles, given by
\begin{align}
\nonumber&\frac{D^2\eta^{\hat{r}}}{D\tau^2}=\frac{1}{\left(r^4+2l^2Mr+Q^2l^2\right)^3}\left[6Q^4\left(l^4Mr+2l^2r^4\right)-Q^6l^4\right.\\
\nonumber &+2Mr^3\left(4l^4M^2-14l^2Mr^3+r^6\right)\\
& \label{TF_r} -3Q^2r^2\left(r^6
\left.-4l^4M^2+4l^2Mr^3\right)\right]\,\eta^{\hat{r}},\\
\label{TF_t}&\frac{D^2\eta^{\hat{\theta}}}{D\tau^2}=\frac{\left[Q^2\left(2l^2Mr+r^4\right)-Q^4l^2-Mr^2\left(r^3-4l^2M\right)\right]}{\left(r^4+2l^2Mr+Q^2l^2\right)^2}\,\eta^{\hat{\theta}},\\
\label{TF_p}&\frac{D^2\eta^{\hat{\phi}}}{D\tau^2}=\frac{\left[Q^2\left(2l^2Mr+r^4\right)-Q^4l^2-Mr^2\left(r^3-4l^2M\right)\right]}{\left(r^4+2l^2Mr+Q^2l^2\right)^2}\,\eta^{\hat{\phi}}.
\end{align}
From Eqs.~\eqref{TF_r}-\eqref{TF_p}, we see how the tidal forces for the charged Hayward BH spacetime depend on the parameters $M$, $Q$ and $l$. For $l=0$, Eqs.~\eqref{TF_r}-\eqref{TF_p} reduce to
\begin{align}
&\left.\frac{D^2\eta^{\hat{r}}}{D\tau^2}\right|_{l=0}=\left(\frac{2\,M}{r^3}-\frac{3\,Q^2}{r^4}\right)\eta^{\hat{r}},\\
&\left.\frac{D^2\eta^{\hat{\theta}}}{D\tau^2}\right|_{l=0}=\left(-\frac{M}{r^3}+\frac{Q^2}{r^4}\right)\eta^{\hat{\theta}},\\
&\left.\frac{D^2\eta^{\hat{\phi}}}{D\tau^2}\right|_{l=0}=\left(-\frac{M}{r^3}+\frac{Q^2}{r^4}\right)\eta^{\hat{\phi}},
\end{align}
which are the tidal forces for the RN spacetime~\cite{RN_TF}. Moreover, for $Q=0$, we have
\begin{align}
&\left.\frac{D^2\eta^{\hat{r}}}{D\tau^2}\right|_{Q=0}=2Mr^3\frac{\left(r^6-14l^2Mr^3+4l^4M^2\right)}{\left(r^4+2l^2Mr\right)^3}\,\eta^{\hat{r}},\\
&\left.\frac{D^2\eta^{\hat{\theta}}}{D\tau^2}\right|_{Q=0}=Mr^2\frac{\left(4l^2M-r^3\right)}{\left(r^4+2l^2Mr\right)^2}\eta^{\hat{\theta}},\\
&\left.\frac{D^2\eta^{\hat{\phi}}}{D\tau^2}\right|_{Q=0}=Mr^2\frac{\left(4l^2M-r^3\right)}{\left(r^4+2l^2Mr\right)^2}\eta^{\hat{\phi}},
\end{align}
which are the tidal forces for the regular chargeless Hayward BH spacetime.
We see that the tidal forces in radial and angular directions may vanish, in contrast with the Schwarzschild case \cite{dinverno, hobson}. In the next subsections, we study in details the tidal forces in the charged Hayward BH spacetime.
\subsection{Radial tidal force}
\label{RTF_sec}
From Eq.~\eqref{TF_r}, we see that the tidal force on the radial direction remains finite at $r=0$, being equal to 
\begin{align}
\label{TF_r0}\lim_{r\rightarrow 0}&\frac{D^2\eta^{\hat{r}}}{D\tau^2}=-\frac{1}{l^2}\,\eta^{\hat{r}}.
\end{align}
This result contrasts with the Schwarzschild, for which the radial tidal force is infinity at $r=0$, where the singularity is located. We note that a radially infalling particle does not reach $r=0$, because it bounces back at $r=R_{\text{stop}}$. Moreover, from Eq.~\eqref{TF_r} we obtain that the tidal force in the radial direction may vanish at two different values of the radial coordinate. The vanishing radial tidal force occurs when the numerator of Eq.~\eqref{TF_r} is zero, resulting in a polynomial equation of ninth degree.

\begin{figure}[p]
  \centering
  \subfigure{\includegraphics[scale=0.73]{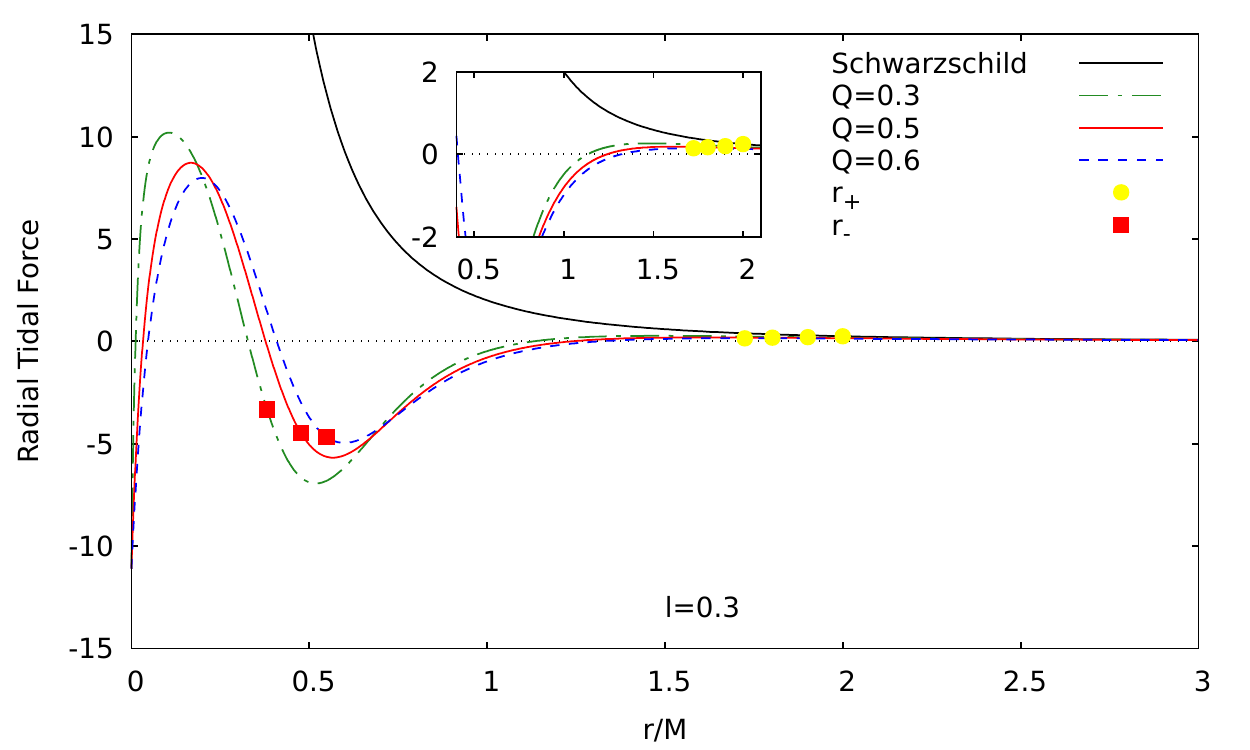}}
  \\
\subfigure{\includegraphics[scale=0.73]{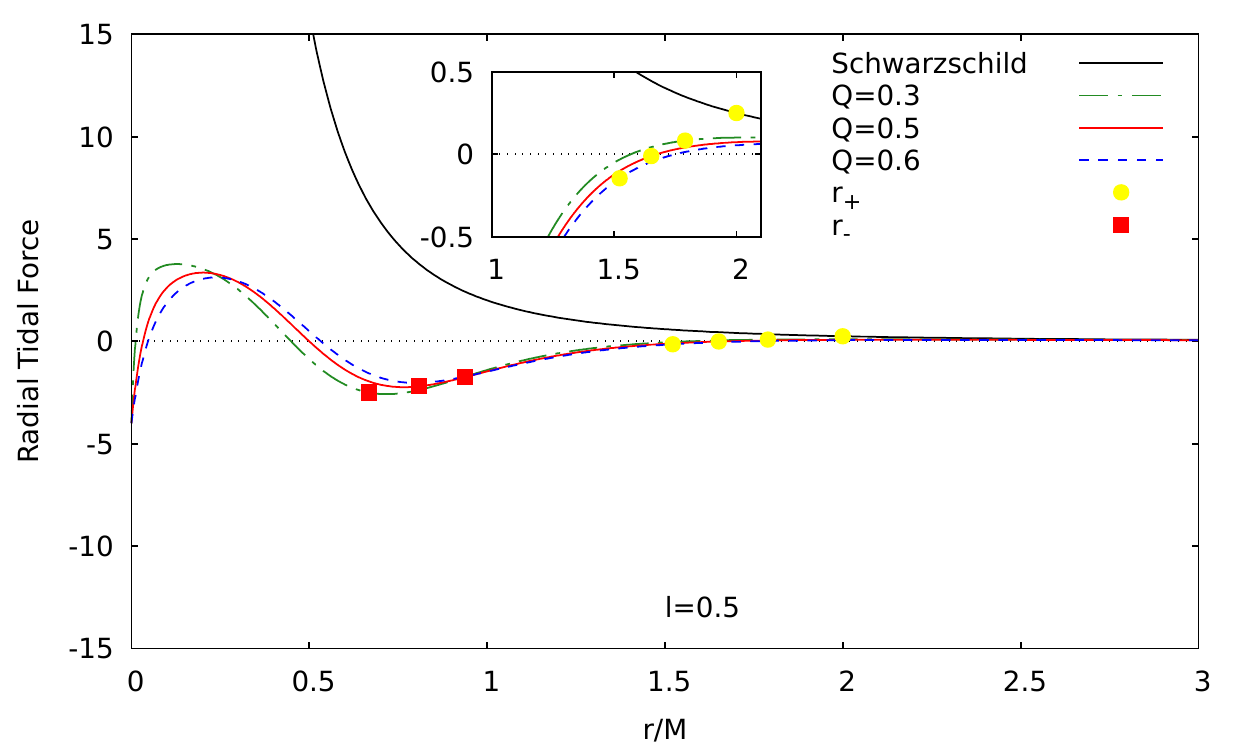}}\\
  \subfigure{\includegraphics[scale=0.73]{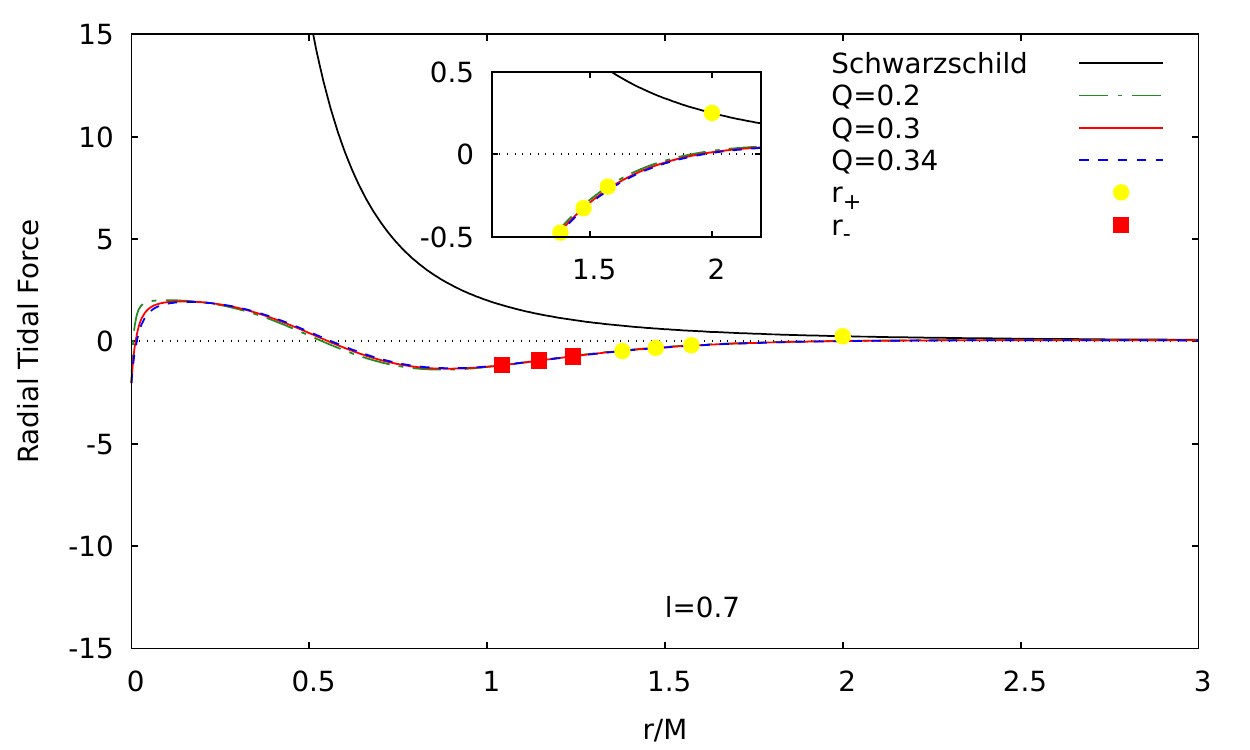}}
\caption{Radial tidal force in the charged Hayward BH spacetime, as a function of the radial coordinate $r$. We note that the radial tidal force may vanish at two different points, in contrast to what happens in the Schwarzschild spacetime. (For comparison, we also plot the radial tidal force for the Schwarzschild spacetime.) The value of the tidal force at $r=0$ is given by Eq.~\eqref{TF_r0}. We have set $b=100M$. }
\label{RTF}
\end{figure}

The radial tidal force as a function of the radial coordinate in charged Hayward BH spacetime is plotted in Fig.~\ref{RTF}, for different values of $Q$ and $l$. We see that the radial tidal force may vanish and change sign. The more external point ($r=R_0^{\ rft}$) where the radial tidal force vanishes can be located outside the event horizon, as can be seen in Fig.~\ref{R_stop_Fig}. Therefore this effect may, in principle, be observable. Vanishing radial tidal forces do not occur in Schwarzschild spacetime, but they occur in RN spacetime \cite{RN_TF}, as well as in other regular BH spacetimes \cite{RBH_TF}. 
\afterpage{\clearpage}

\begin{figure}
  \centering
  \subfigure{\includegraphics[scale=0.73]{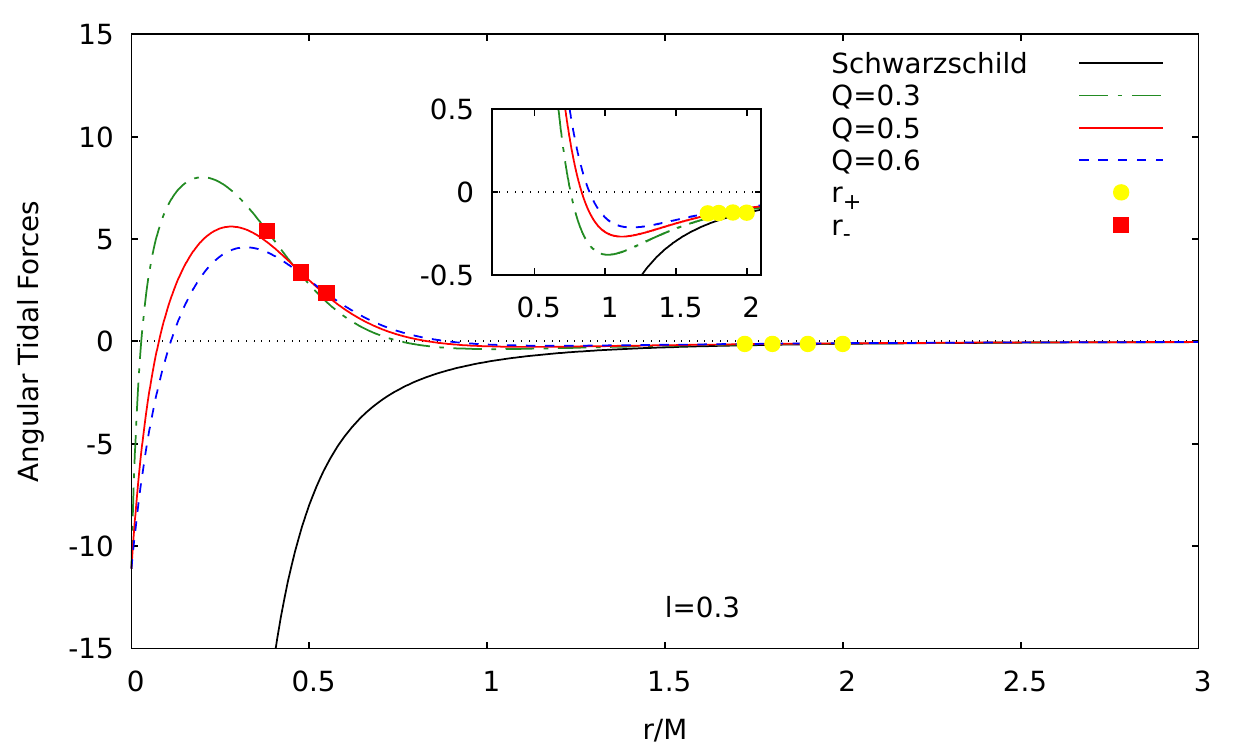}}
  \\
\subfigure{\includegraphics[scale=0.73]{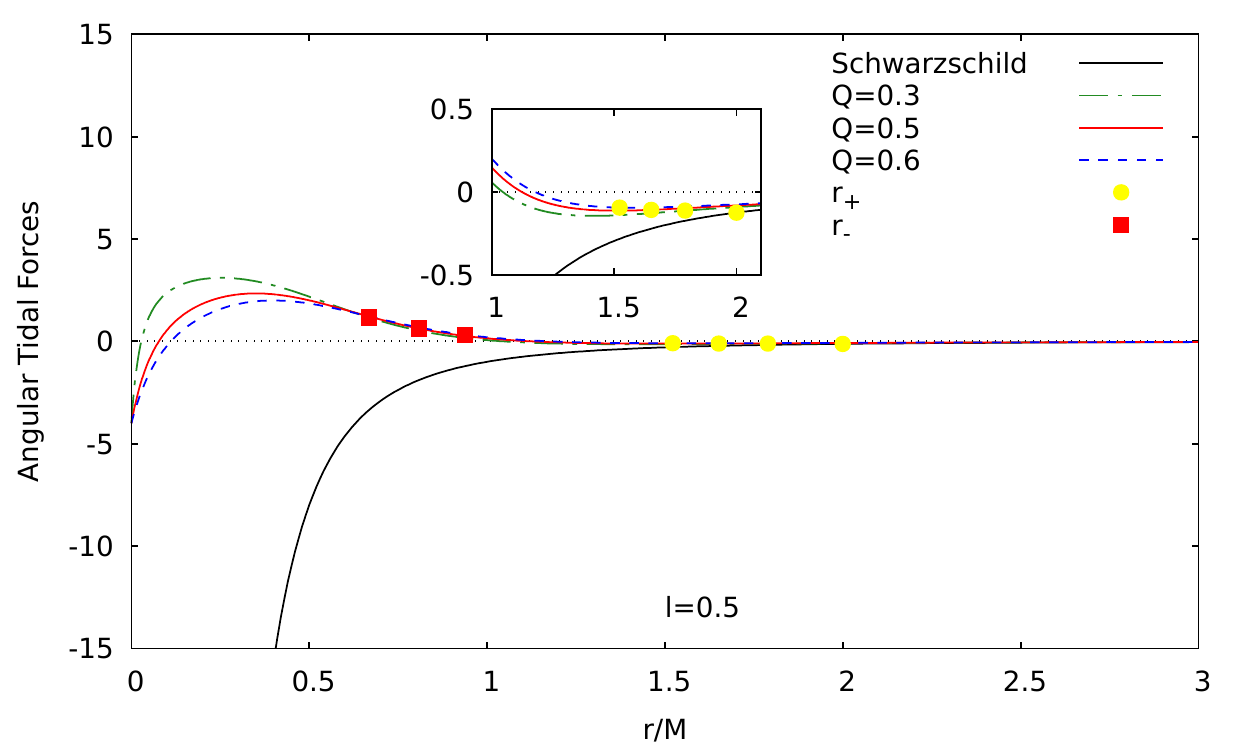}}\\
  \subfigure{\includegraphics[scale=0.73]{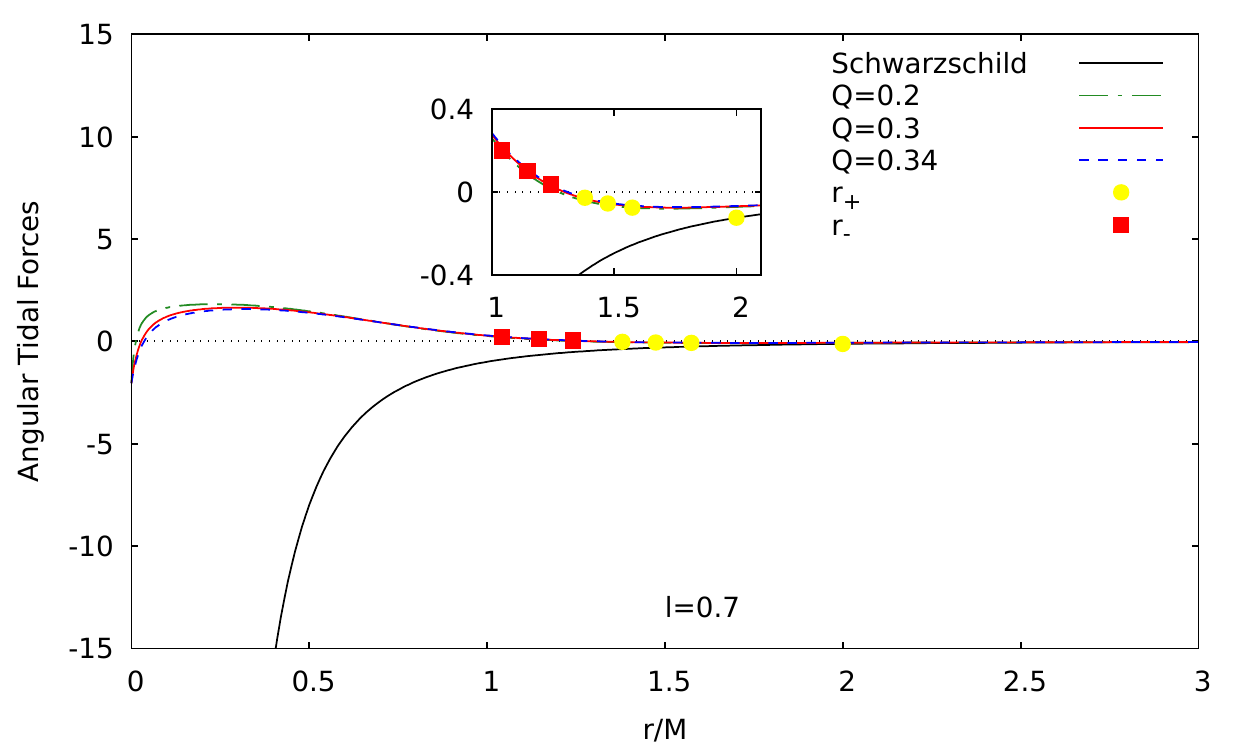}}
\caption{Angular tidal forces as a function of the radial coordinate $r$. We note that the radial tidal forces of the charged Hayward BH spacetime may vanish, in contrast to what happens in the Schwarzschild case. The value of the tidal force at $r=0$ is given by Eq.~\eqref{ATF_r0}. We have set $b=100M$.}
\label{ATF}
\end{figure}

\afterpage{\clearpage}
\begin{figure}
  \centering
  \subfigure{\includegraphics[scale=0.73]{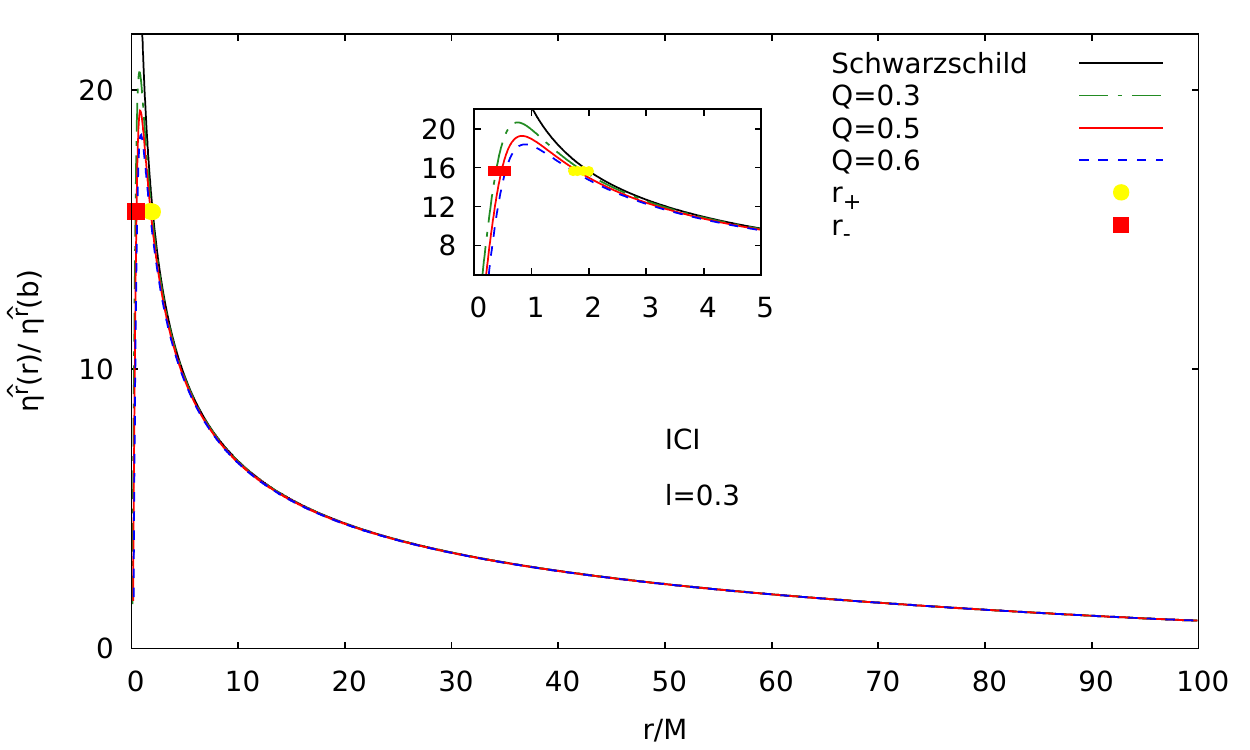}}
  \\
\subfigure{\includegraphics[scale=0.73]{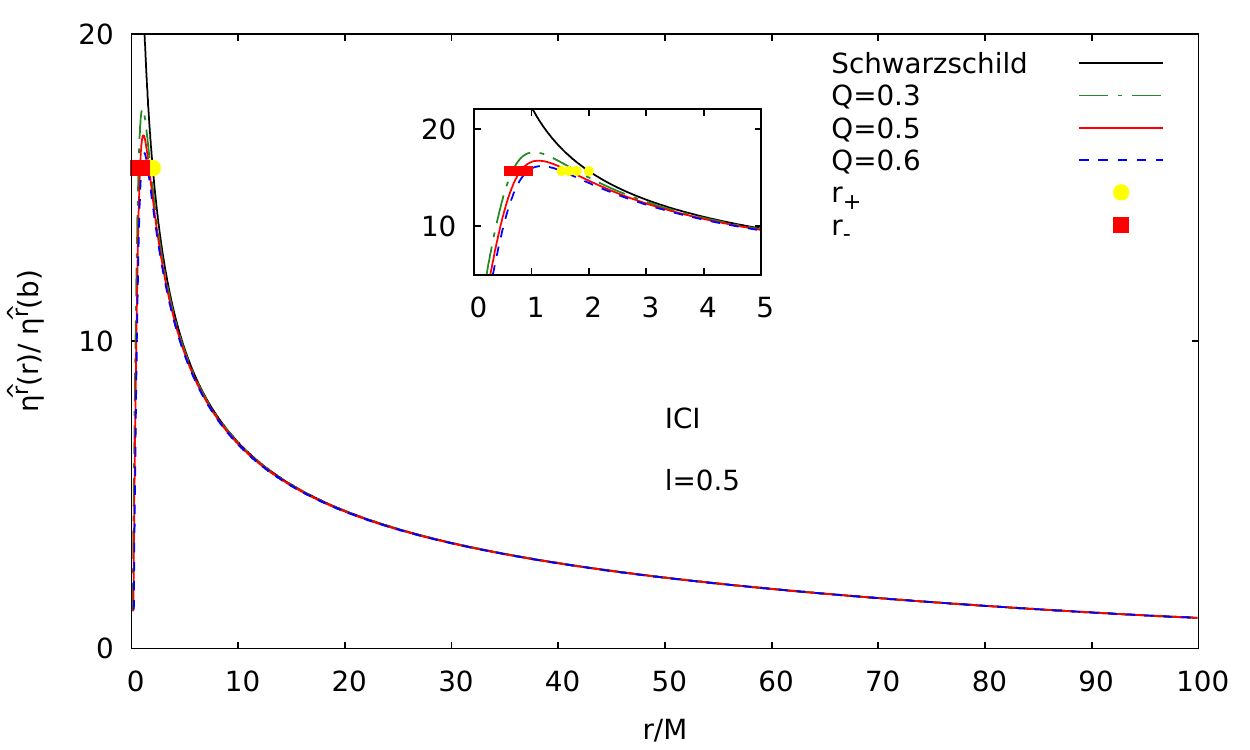}}\\
  \subfigure{\includegraphics[scale=0.73]{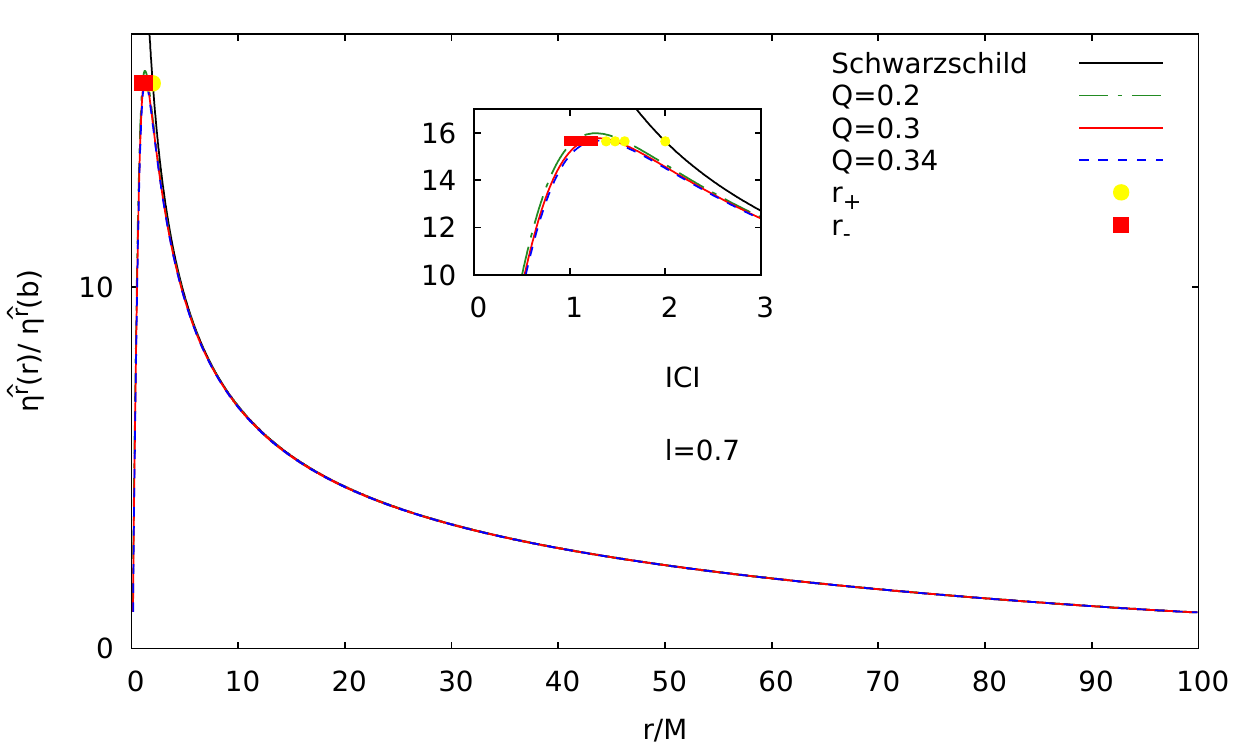}}
\caption{Radial component of the geodesic deviation vector of charged Hayward BHs with ICI, as a function of $r$, and for different values of $Q$  and $l$. In We have chosen $b=100M$. For comparison, we also plot the radial component of the geodesic deviation vector in the Schwarzschild case.}
\label{DVR_IC1}
\end{figure}

\begin{figure}
  \centering
  \subfigure{\includegraphics[scale=0.73]{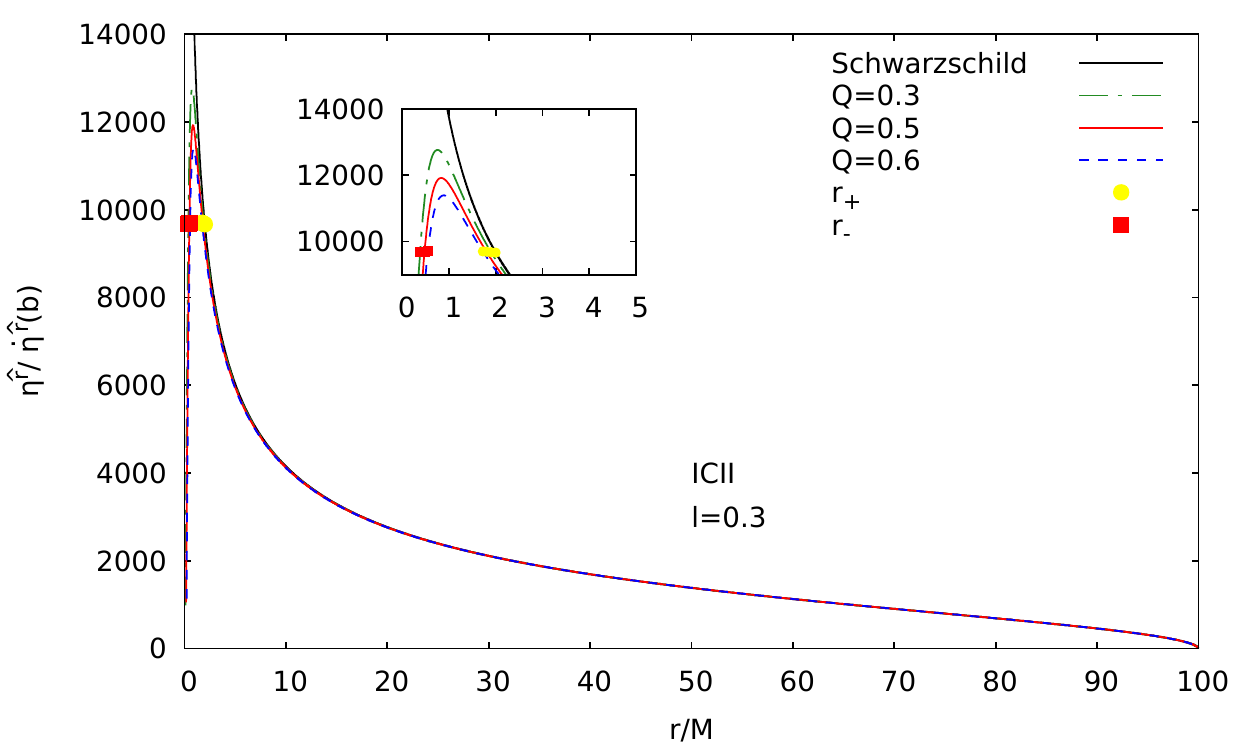}}
  \\
\subfigure{\includegraphics[scale=0.73]{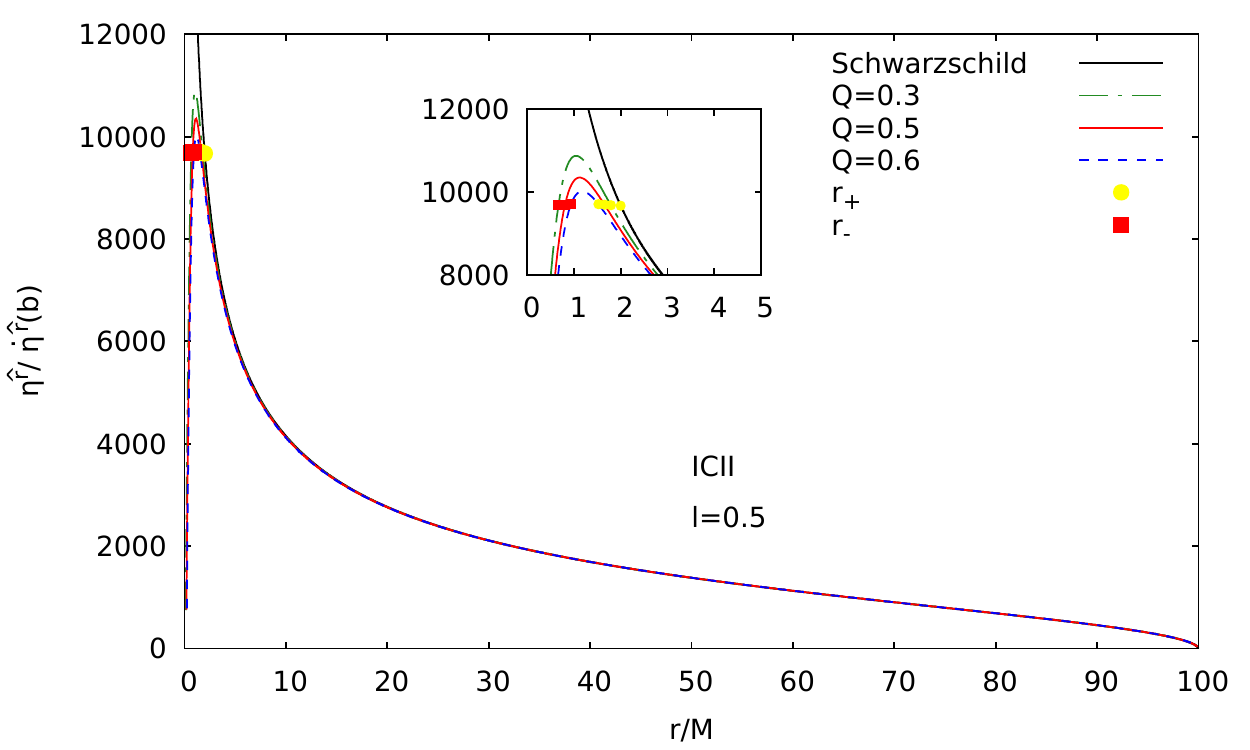}}\\
  \subfigure{\includegraphics[scale=0.73]{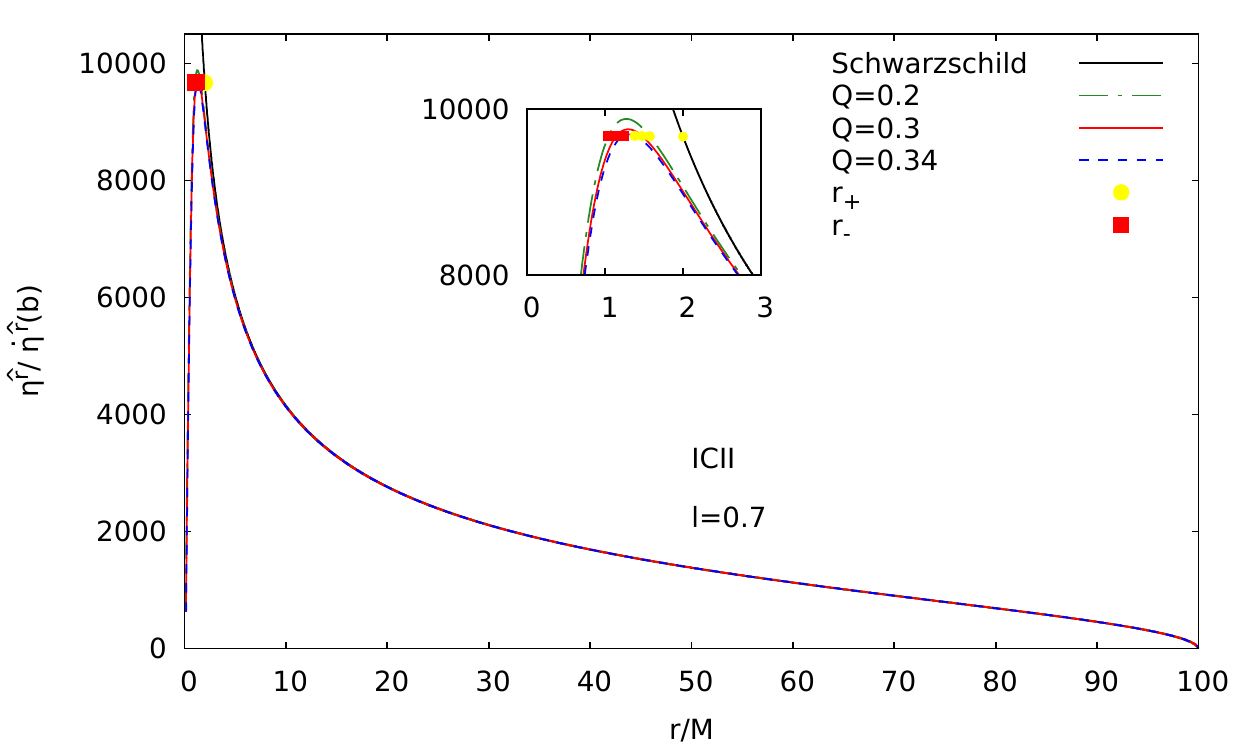}}
\caption{As in Fig.~\ref{DVR_IC1}, we plot the radial component of the geodesic deviation vector as a function of $r$, now with ICII (and $b=100M$), for different values of $Q$  and $l$.}
\label{DVR_IC2}
\end{figure}

\afterpage{\clearpage}
\begin{figure}
  \centering
  \subfigure{\includegraphics[scale=0.73]{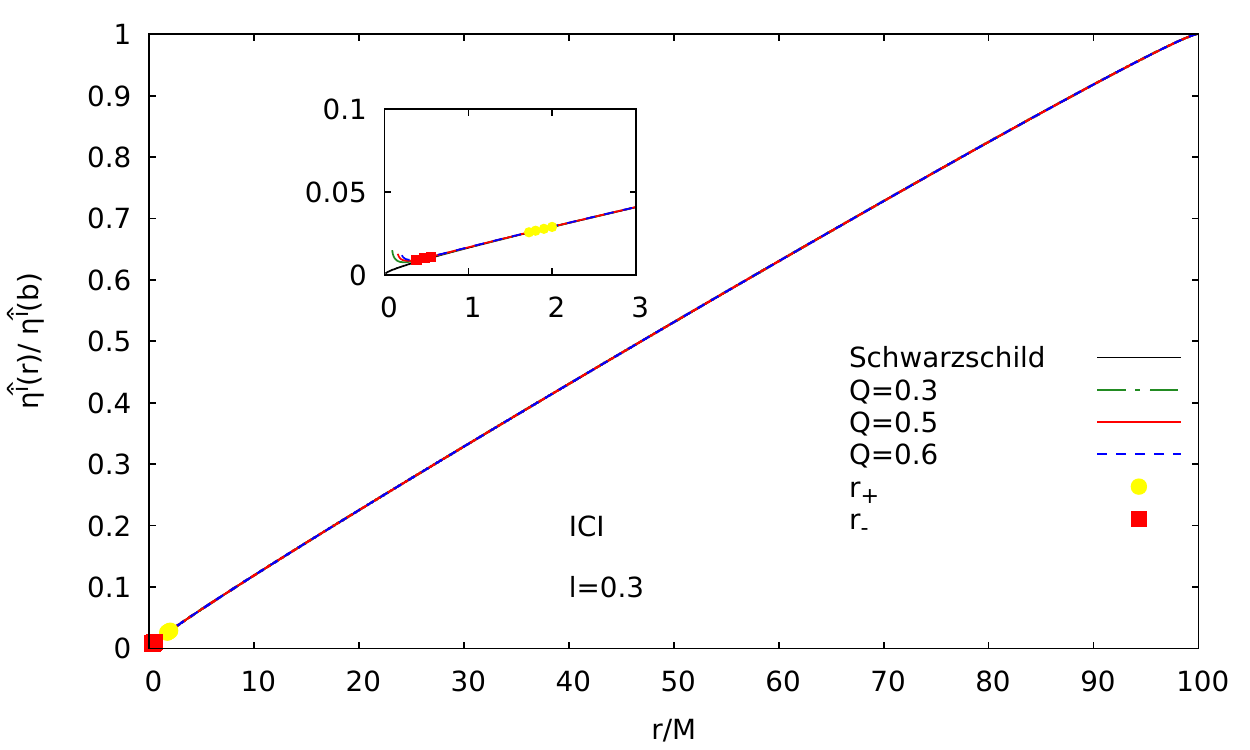}}
  \\
\subfigure{\includegraphics[scale=0.73]{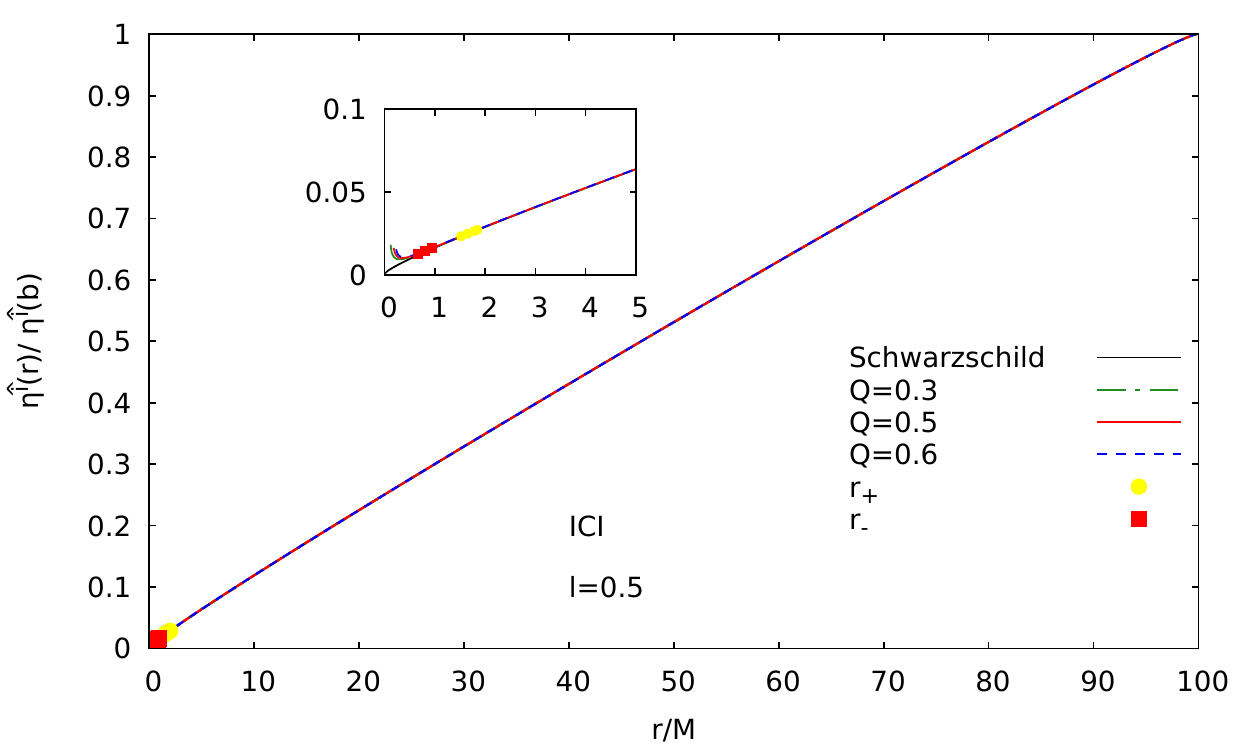}}\\
  \subfigure{\includegraphics[scale=0.73]{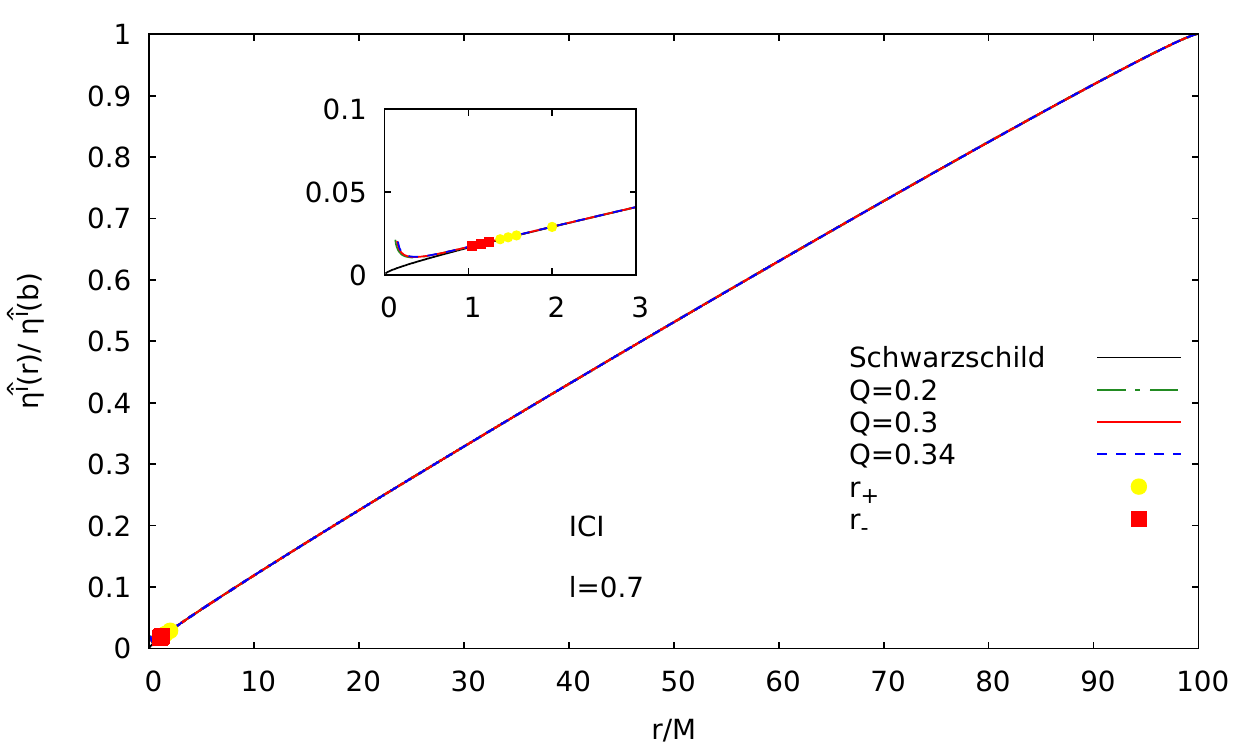}}
\caption{Angular components of the geodesic deviation vector of charged Hayward BHs, as a function of $r$, with ICI, $b=100M$ and for different values of Q and l. The angular components of the geodesic deviation vector for the Schwarzschild case are also shown.}
\label{DVi_IC1}
\end{figure}
\begin{figure}
  \centering
  \subfigure{\includegraphics[scale=0.73]{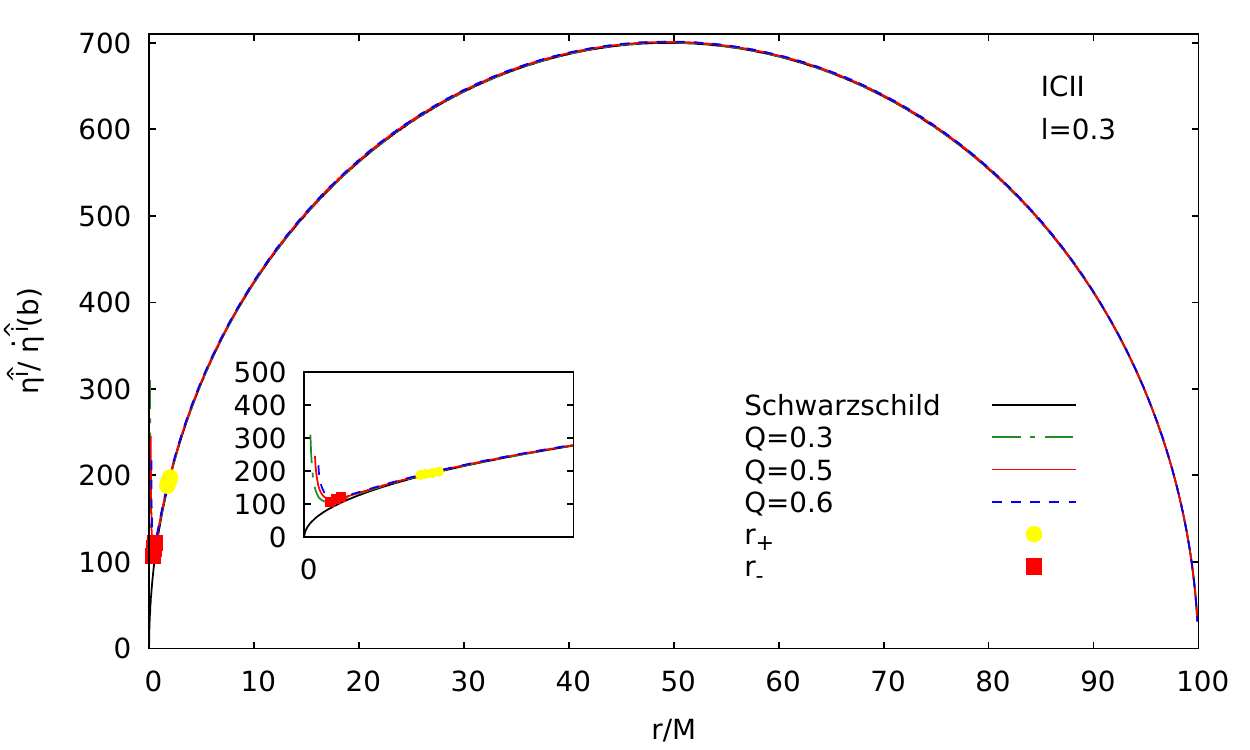}}
  \\
\subfigure{\includegraphics[scale=0.73]{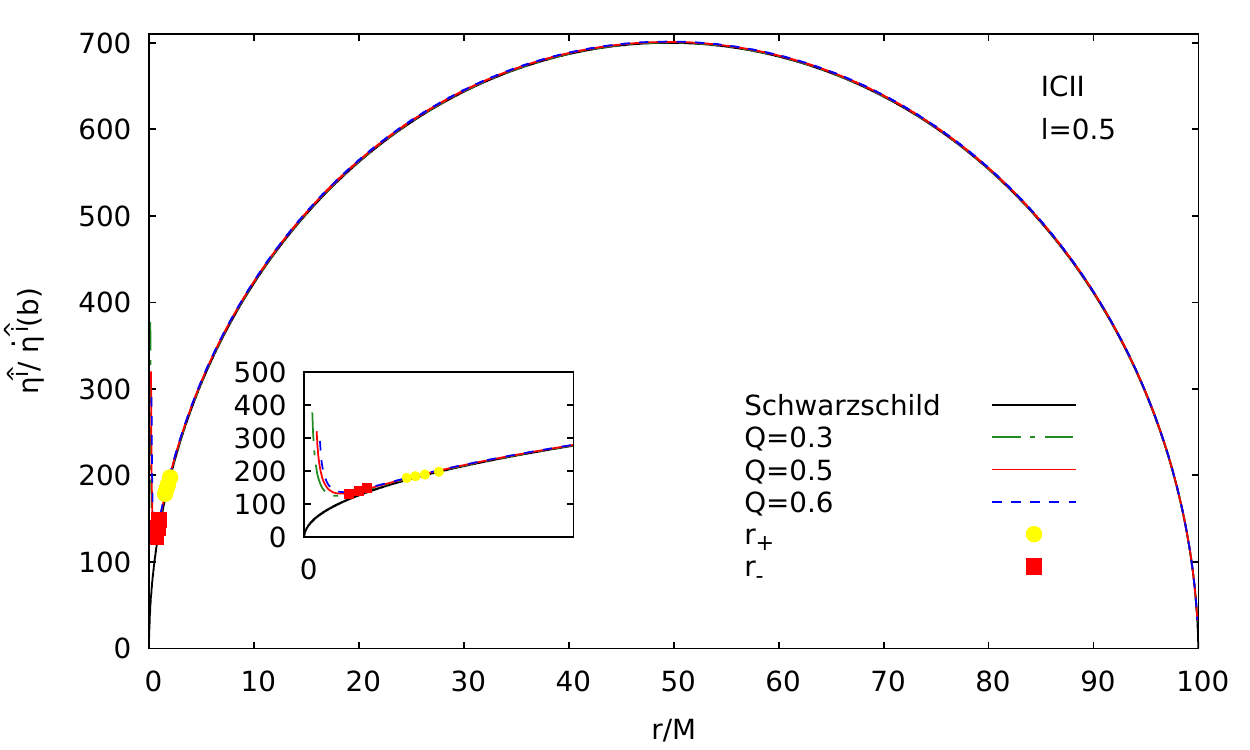}}\\
  \subfigure{\includegraphics[scale=0.73]{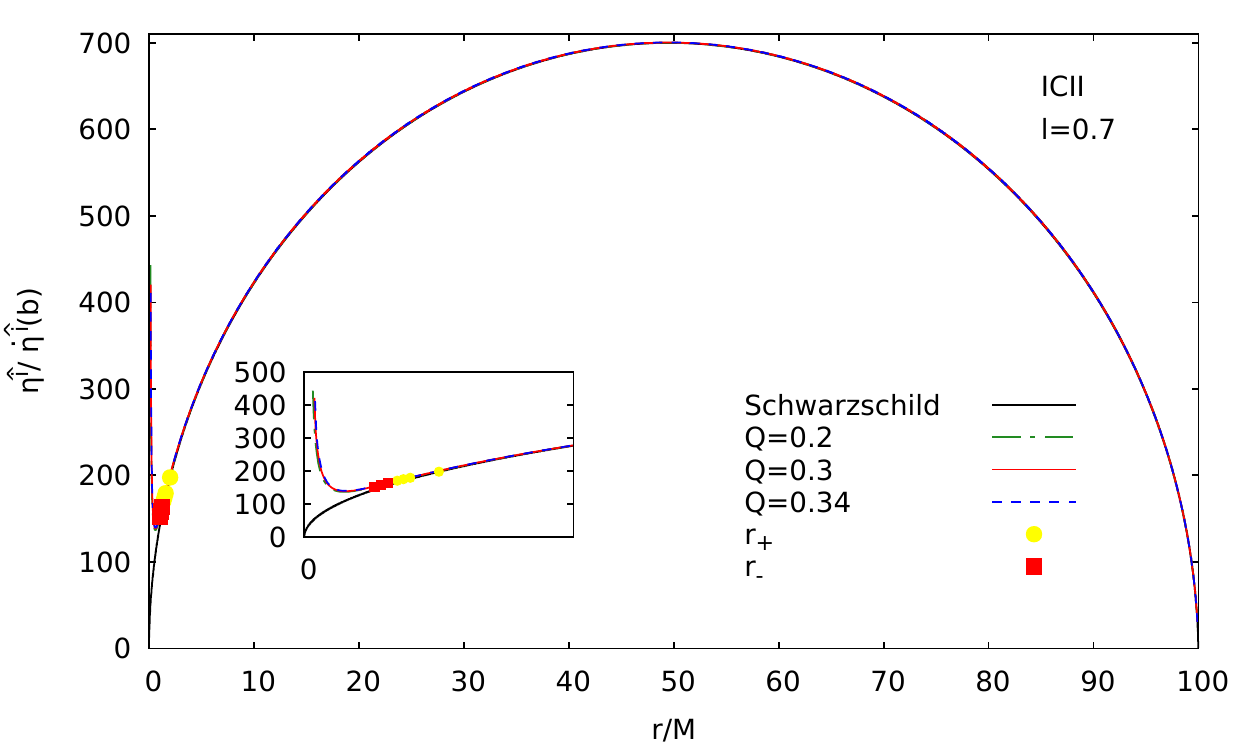}}
\caption{As in Fig.~\ref{DVi_IC1}, we plot the angular components of the geodesic deviation vector as a function of $r$, now with ICII (and $b=100M$), and for different values of Q and l.}
\label{DVi_IC2}
\end{figure}

\subsection{Angular tidal forces}
Similarly to the radial tidal forces, the angular tidal forces remain finite at $r=0$. From Eqs. \eqref{TF_t}-\eqref{TF_p}, we obtain
\begin{align}
\label{ATF_r0}\lim_{r\rightarrow 0}&\frac{D^2\eta^{\hat{i}}}{D\tau^2}=-\frac{1}{l^2}\,\eta^{\hat{i}},
\end{align}
where $i=(\theta,\phi)$. This result is also in contrast with the Schwarzschild, for which the angular tidal forces diverge at $r=0$.
In Fig.~\ref{ATF}, we plot the angular tidal forces as a function of the radial coordinate, for different values of $Q$ and $l$. We note that the angular tidal forces may vanish. The vanishing tidal force point ($r=R_0^{\ atf}$) is located outside the Cauchy horizon and inside the event horizon, as  it can be seen in Fig.~\ref{R_stop_Fig}.

\section{Geodesic deviation vector}
\label{dev_vec_sol}
Let us analyze the geodesic deviation vector that describes the deformation, due to tidal forces, of a body constituted of chargeless dust infalling radially in a charged Hayward BH. To solve Eqs.~\eqref{TF_r}-\eqref{TF_p}, we start by rewriting them as
\begin{align}
\label{TF_r2}&\frac{D^2\eta^{\hat{r}}}{D\tau^2}=-\frac{f''}{2}\eta^{\hat{r}},\\
\label{TF_t2}&\frac{D^2\eta^{\hat{i}}}{D\tau^2}=-\frac{f'}{2\,r}\,\eta^{\hat{i}},
\end{align}
where the primes denote differentiation with respect to the radial coordinate. We can write Eqs.~\eqref{TF_r2}-\eqref{TF_t2} in terms of derivatives with respect to the radial coordinate, by using that $dr/d\tau=-\sqrt{E^2-f(r)}$. The result is
\begin{align}
\label{geo_devr}&\left(E^2-f(r)\right)\eta^{\hat{r}''}-\frac{f(r)'}{2}\eta^{\hat{r}'}+\frac{f(r)''}{2}\,\eta^{\hat{r}}=0,\\
\label{geo_devi}&\left(E^2-f(r)\right)\eta^{\hat{i}''}-\frac{f(r)'}{2}\eta^{\hat{i}'}+\frac{f(r)'}{2\,r}\,\eta^{\hat{i}}=0.
\end{align}
As pointed out in Ref.~\cite{RN_TF}, the components of the geodesic deviation vector, solutions of Eqs.~\eqref{geo_devr}-\eqref{geo_devi}, are
\begin{align}
\label{eq_etar}&\eta^{\hat{r}}(r)=\sqrt{E^2-f}\left[C_1+C_2\int\frac{dr}{\left(E^2-f\right)^{3/2}}\right],\\
\label{eq_etai}&\eta^{\hat{i}}(r)=\left[C_3+C_4\int\frac{dr}{r^2\left(E^2-f\right)^{1/2}}\right]\,r,
\end{align}
where $C_1$, $C_2$, $C_3$ and $C_4$ are integration constants. In order to determine such integration constants, we choose two different types of initial conditions:
\begin{align}
\label{IC_I}&\eta^{\hat{\alpha}}(b)>0,\ \dot{\eta}^{\hat{\alpha}}(b)=0, \ \ \ \ \text{(ICI),}\\
\label{IC_II}&\eta^{\hat{\alpha}}(b)=0,\ \dot{\eta}^{\hat{\alpha}}(b)>0, \ \ \ \ \text{(ICII),}
\end{align} 
with $b>r_+$. The ICI corresponds to release from rest, at the radial coordinate $r=b$, a body constituted of dust with no internal motion. On the other hand, ICII corresponds to let a body constituted of dust to ``explode'' at $r=b$. In the next subsections, we study in details the radial and angular components of the geodesic deviation vector for ICI and ICII.

\subsection{Radial component of the deviation vector}

In Fig.~\ref{DVR_IC1} we plot the geodesic deviation vector in the radial direction, obtained solving Eq.~\eqref{geo_devr} with ICI. We note that  far away from the BH the radial component of the geodesic deviation vector is essentially the same for different values of $Q$ and $l$. We also note that during the radial infall, starting at $r=b$, the radial geodesic deviation vector always increases outside the event horizon. It reaches a maximum between the event horizon and the Cauchy horizon, and then decreases until $R_{\text{stop}}$. This reflects the change in the sign of the tidal force in the radial direction.

In Fig.~\ref{DVR_IC2} we plot the geodesic deviation vector in the radial direction, obtained solving Eq.~\eqref{geo_devr} with ICII.  We note that during the radial infall the behavior of the radial component of the geodesic deviation vector subjected to ICII is qualitatively similar to ICI case.

For comparison, we also show in Figs.~\ref{DVR_IC1} and \ref{DVR_IC2} the radial component of the geodesic deviation vector for the Schwarzschild spacetime, and we note that it goes to infinity at the origin ($r=0$). This reflects the infinite stretching tidal force  in the radial direction at the singularity of the Schwarzschild spacetime.

\subsection{Angular components of the deviation vector}

In Fig.~\ref{DVi_IC1} we plot the angular components of the geodesic deviation vector, obtained solving Eq.~\eqref{geo_devi}, subjected to ICI. We note that for $r\gg r_+$, the angular components of the geodesic deviation vector have similar behavior regardless of the electric charge $Q$. This happens due to the fact that for large values of $r$ the spacetime looks similar for different values of $Q$. We also note that the angular components of the geodesic deviation vector decrease linearly with $r$, as expected, since all individual geodesic with $L=0$ is radial.

In Fig.~\ref{DVi_IC2} we plot the components of the geodesic deviation vector in the angular directions, obtained solving Eq.~\eqref{geo_devi} with ICII. We note that the angular components of the geodesic deviation vector initially increase, reach a maximum around $b/2$ and then start decreasing. This is an effect of the compressing tidal forces in angular directions. The angular components of the geodesic deviation vector reach a minimum inside the event horizon and then increase again. This reflects the change in the sign of the tidal forces in the angular directions.

We also show in Figs.~\ref{DVi_IC1} and \ref{DVi_IC2} the angular components of the geodesic deviation vector for the Schwarzschild spacetime, and we note that it goes to zero at the origin ($r=0$). This reflects the infinite compressing tidal forces in the angular directions at the singularity of the Schwarzschild spacetime.

\section{Conclusion}
\label{Conclusion}
 We studied in details the tidal forces in the electrically charged Hayward BH spacetime. We noted that tidal forces in such spacetime are finite at the origin of the radial coordinate, similarly to the electrically uncharged case. We have shown that the expressions for tidal forces in RN and  chargeless Hayward BH spacetimes are recovered by taking the suitable limits for the parameters present in the line element.
 
 We note that both radial and angular components of the tidal force may vanish. Therefore, the  tidal force in the radial direction may become compressing instead of stretching, and the tidal forces in the angular directions may become stretching instead of compressing. The radial tidal force may vanish outside the event horizon, depending on the  values of the electric charge and the parameter $l$. Therefore, this effect can, in principle, be observable for some charged Hayward BHs. Concerning the angular tidal forces, they may vanish only inside the event horizon. This behavior is different from the case of Schwarzschild spacetime (in which tidal forces do not vanish) but is qualitatively similar to the RN spacetime.
 
We have found the radial and angular components of the geodesic deviation vector. Two types of initial condition were studied. One initial condition is related to releasing from rest, far from the BH, a body constituted of dust with no internal motion. The other initial condition is related to a body constituted of dust ``exploding'' far from the BH. We note that the geodesic deviation vector in charged Hayward BHs can be quite distinctive from the Schwarzschild BH. For instance, the radial component of the geodesic deviation vector in the charged Hayward BHs is finite next to the origin of the radial coordinate, while in the Schwarzschild case it tends to infinity. The angular components of the geodesic deviation vector in the charged Hayward BHs remain greater than zero next to the origin of the radial coordinate, while for the Schwarzschild case it goes to zero.


\section*{Acknowledgements}
We thank A. Higuchi and C. Herdeiro for useful discussions.
This research was financed in part by Coordena\c{c}\~ao de Aperfei\c{c}oamento de Pessoal de N\'ivel Superior (CAPES, Brazil) -- Finance Code 001, and by Conselho Nacional de Desenvolvimento Cient\'ifico e Tecnol\'ogico (CNPq, Brazil). This research has also received funding from the European Union's Horizon 2020 research and innovation programme under the H2020-MSCA-RISE-2017 Grant No. FunFiCO-777740.

\end{document}